\documentclass[aps,prd,onecolumn,groupedaddress,nofootinbib,longbibliography,superscriptaddress]{revtex4}

\usepackage{amsfonts,amsmath,amssymb}
\usepackage{mathrsfs}
\usepackage{nicefrac}
\usepackage{graphicx}
\usepackage{color}
\usepackage{soul} 
\usepackage{hyperref}
\usepackage{accents}
\usepackage{subfigure}
\usepackage[normalem]{ulem}
\usepackage{hyperref}
\usepackage{xcolor}
\usepackage{fleqn}

\hypersetup{
    colorlinks=true, 
    linktoc=page,    
    linkcolor=blue,  
    urlcolor=magenta
}

\newcommand{\be}{\nopagebreak[3]\begin{equation}}
\newcommand{\ee}{\end{equation}}

\newcommand{\ba}{\nopagebreak[3]\begin{eqnarray}}
\newcommand{\ea}{\end{eqnarray}}

\begin{document}
\title{Echoes in the Kerr/CFT correspondence}

\author{Ramit Dey}
\email[]{ramitdey@gmail.com}
\affiliation{School of Mathematical and Computational Sciences, Indian Association for the Cultivation of Science, 
Kolkata-700032, India}
\author{Niayesh Afshordi}
\email[]{nafshordi@pitp.ca }
\affiliation{Department of Physics and Astronomy, University of Waterloo,
200 University Ave W, N2L 3G1, Waterloo, Canada
}
\affiliation{Waterloo Centre for Astrophysics, University of Waterloo, Waterloo, ON, N2L 3G1, Canada}
\affiliation{Perimeter Institute For Theoretical Physics, 31 Caroline St N, Waterloo, Canada}

\begin{abstract}
 The  Kerr/CFT correspondence is a possible route to gain insight into the quantum theory of gravity in the near-horizon region of a Kerr black hole via a  dual holographic conformal field theory (CFT).  Predictions of the black hole entropy, scattering cross-section and the quasi normal modes from the dual holographic CFT corroborate this proposed correspondence. More recently, it has been suggested that quantum gravitational effects in the near-horizon region of a black hole may drastically modify the classical general relativistic description, leading to potential observable consequences.
In this paper, we study the absorption cross-section and quasi normal modes of a horizonless Kerr-like exotic compact object (ECO) in the dual CFT picture. Our analysis suggests that the near-horizon quantum modifications of the black hole can be understood as finite size and/or finite $N$ effects in the dual CFT. Signature of the near-horizon modification to a black hole geometry manifests itself as delayed {\it echoes} in the ringdown (i.e. the postmerger phase) of a binary black hole coalescence. From our dual CFT analysis we show how the length of the circle, on which the dual CFT lives, must be related to the echo time-delay that depends on the position of the near-horizon quantum structure.  We further derive the reflectivity of the ECO membrane in terms of the CFT modular parameters, showing that it takes the Boltzmann form.  
\end{abstract}

\maketitle
\section{Introduction}
Over the past few decades, the quasi normal modes (QNMs) of various black holes and exotic compact objects (ECOs) have been studied extensively \cite{Kokkotas:1999bd,Berti:2009kk}. The ringdown in the postmerger phase of a binary coalescence can be described in terms of a superposition of the QNM. This is sensitive to any modification to the near-horizon physics of the compact object that can modify/remove the horizon, as the boundary condition imposed at the horizon for obtaining the QNM spectrum, depends on the nature/presence of the horizon. Precise detection of the QNM by analysing the ringdown spectrum can give us a fair idea about the exact nature of the black hole (or ECO) horizon and test the accuracy of general relativity in the strong gravity regime. 
From the ringdown signal, it is  possible to probe some of the quantum aspects of a black hole and analyse  key signatures of quantum gravity leading to near-horizon modification of the black hole geometry. Such near-horizon modifications of the black hole are usually realised in models (Firewalls \cite{Almheiri:2012rt}, fuzzball geometry \cite{Mathur:2005zp}, gravastars \cite{Mazur:2001fv}, 2-2 holes \cite{Holdom:2016nek})  resolving the information loss paradox or stems from the modified dynamics in the strong gravity regime addressing the dark energy problem \cite{PrescodWeinstein:2009mp}.

Echoes in the ringdown part of the gravitational wave spectrum \cite{Cardoso:2016rao,Cardoso:2016oxy,Abedi:2016hgu,Abedi:2020ujo,Oshita:2020dox} are considered as a signature of quantum modifications to the classical black hole horizon and reports of tentative detection of these echoes, \cite{abedi2018echoes,Abedi:2016hgu,Holdom:2019bdv} makes the subject even more interesting. Attempts to provide a microscopic description of the spectrum of quantum black holes include a quantum multilevel system \cite{Wang:2019rcf,Oshita:2019sat}, quantum corrected black holes in the braneworld scenario \cite{Dey:2020lhq}, interpreting the echo time in terms of the scrambling time \cite{Saraswat:2019npa}, black hole area quantization \cite{Cardoso:2019apo,Coates:2019bun}, or 2-2 holes in asymptotically free Quadratic Quantum Gravity \cite{Holdom:2016nek,Conklin:2017lwb}. Removal of the black hole horizon due quantum effects in the near-horizon region (e.g., through introducing a partially reflective membrane in front of the would-be horizon)  would modify the quasi normal modes and the QNM spectrum of these horizonless ECOs are different from that of the classical  black holes as a purely ingoing boundary condition at the horizon cannot be imposed for these horizonless compact objects. Since the motivation for near-horizon modifications to the black hole geometry comes from quantum gravitational effects, in principle it must be possible to explain the removal/modification of the event horizon from the near-horizon degrees of freedom within a theory of quantum gravity.

The Kerr/CFT correspondence conjectures that quantum gravity in the near-horizon region of the Kerr black hole is dual to a two-dimensional thermal CFT
\cite{Guica:2008mu,Compere:2012jk}. The correspondence \cite{Guica:2008mu} was originally shown to exist for an extremal Kerr black holes and it relied heavily on the decoupling of the near-horizon region (NHEK) with an exact $SL(2,R)\times U(1)$ symmetry. The $SL(2,R)$ symmetry governed the behaviour of the near-horizon scattering cross-section and it was explicitly shown that the dual two-dimensional chiral CFT gives the same results as the gravity computation. 
For non-extremal Kerr black holes, the realisation of the correspondence was less trivial but it was shown  that there is an underlying hidden $SL(2,R)\times SL(2,R)$ present in the dynamics of a probe scalar field close to the horizon (in the low frequency limit)\cite{Castro:2010fd}.
Based on this ``hidden conformal symmetry'', and the dual  two-dimensional CFT living in the near-horizon region of the Kerr black hole,  the  black hole entropy as well as the absorption cross-section of the black hole \cite{Castro:2010fd,Haco:2018ske} was computed. It was further shown that the probe scalar field equation can be written as the $SL(2,R)\times SL(2,R)$ Casimir, establishing the conjectured correspondence even further. 
In this case the local conformal symmetry in the solution space of the wave equation for propagating fields in the Kerr background was the sufficient condition for obtaining the correct scattering cross-section.  In \cite{Haco:2018ske,Haco:2019ggi,Chen:2020nyh} the central charge of the dual CFT was computed using the covariant phase space formalism,  making the correspondence for non-extremal Kerr black holes even stronger. In this paper, we look for a plausible way to  interpret the near-horizon quantum modifications to a Kerr black hole within this holographic setup.

For a Kerr-like ECO with a partially reflective membrane placed in front of the horizon, a probe scalar field would inherit the same hidden conformal symmetry in the near-horizon region as shown in \cite{Castro:2010fd}. 
In this paper, we  study the appropriate holographic CFT description that is dual to the microscopic degrees of freedom in the near-horizon region of such an horizonless ECO. 
We argue that the near-horizon modifications due to quantum effects can originate as finite size effects in the dual field theory, living on a circle of length $L$ (or an Euclidean torus having periodicities of $L$ and $1/T$). For Kerr-like ECOs the reflectivity of the membrane is 
not well understood and it is assumed that an exact quantum gravity computation is required to determine the reflectivity accurately. In this paper, our dual CFT analysis suggests that the reflectivity can be interpreted in terms of a Boltzmann factor which matches with the reflectivity of a quantum horizon as derived in \cite{Oshita:2020dox}.

In the context of AdS/CFT correspondence, attempts were made to understand the bulk geometry dual to a boundary field theory in the strongly coupled regime having a finite volume \cite{Birmingham:2002ph,Solodukhin:2004rv,Solodukhin:2005qy}. Usually, finite-$N$ effects in the boundary theory would correspond to a modification of the semi-classical description of the black hole in the bulk and in particular the black hole horizon is removed (e.g., replaced by a wormhole or a``fractal brick-wall'' \cite{Solodukhin:2005qy}) due to breakdown of the  semi-classical physics close to the horizon at finite Planck length \cite{Kabat:2014kfa}. The brick-wall scenario \cite{tHooft:1984kcu} would be a classic example of such non-perturbative near-horizon modifications where a ``brick-wall'' (or Dirichlet wall) is placed few Planck length away from the horizon to regularize and interpret the black hole entropy as the entropy of particles forming a thermal atmosphere outside the black hole horizon. One must consider the finite-$N$ corrections of the boundary theory to match the spectrum of the bulk  having such a brick-wall within the context of AdS/CFT \cite{Iizuka:2013kma,Kay:2011np,Solodukhin:2005qy}. Since  the exact dual field theory is not known at a microscopic level for the Kerr/CFT correspondence, we do our computation in terms of a generic thermal CFT. 

In this context, perturbation of  the black hole corresponds to perturbation of the dual CFT state with some relevant operator. Thus, to understand the perturbation on the gravity side with a modified boundary condition at the horizon, we must study the thermal two-point function of certain conformal operators.
Usually for a classical black hole,  the dual thermal field theory lives on a toroidal two-manifold and is considered in the high temperature limit  so the effective spatial length of the torus becomes infinite. For the computation of the two-point function in the high temperature limit, a cylindrical approximation of the torus is used which in terms of the spatial cycle ($L$) and temporal cycle ($1/T$) of the torus can be written as $L\gg 1/T$ . However, one must note that the cylindrical limit makes the discrete spectrum of the theory continuous as one might expect for an infinite volume theory. As we will see, the QNM spectrum of the Kerr-like ECO is  discrete, and hence we must consider the dual field theory on a circle of finite length $L$ to establish the duality between the two descriptions. 
We study the  finite-size effects of the covering space (for a dual CFT living on an Euclidean torus having periodicities $L$ and $1/T$) to the thermal two-point function in order to understand the spectrum of the horizonless ECO. As given in \cite{Birmingham:2002ph,Solodukhin:2005qy},
such finite-size effects of the dual CFT can be contrasted with the behaviour of finite $N$ field theories in the strong coupling regime dual to supergravity on $AdS_3$.
Following the original derivation for BTZ black holes \cite{Birmingham:2001pj}, we  show that  the poles of the CFT two-point correlation function on a finite torus match the QNM spectrum of the Kerr-like ECO. The absorption cross-section is  also correctly predicted by the dual CFT  when the length of the circle on which the dual theory lives, is related to the distance of the reflective membrane from the (would-be) horizon, or equivalently the ``echo time-delay"\cite{Wang:2019rcf} for the Kerr black hole/ECO. 

The paper is organized as follows: In section \ref{ECO} we review some of the basic aspects of a Kerr-like ECO and demonstrate how the absorption cross-section and the quasi normal modes differ from the Kerr black hole with a classical event horizon.
In section \ref{hidden}, we discuss the hidden conformal symmetry associated with a probe scalar field in Kerr spacetime and establish the conformal coordinates based on which the Kerr/CFT correspondence is conjectured. We demonstrate how the dual field theory living on a finite circle captures the necessary near-horizon modification of a Kerr-like ECO, finally predicting the  QNM spectrum and absorption cross-section from the dual field theory which matches with the direct bulk computations. In section \ref{observation} we comment on the observational aspects of our findings. Finally, Section \ref{conclusion} summarizes our results and provides some open questions for future study. 

\section{Kerr-like Exotic compact object (ECO)}
\label{ECO}

We consider a model of exotic compact object (ECO) where the exterior spacetime is described  by the Kerr metric but the near-horizon geometry is modified due to the presence of  quantum structures originating from quantum gravitational effects \cite{Maggio:2017ivp,Maggio:2018ivz}. Usually such ECOs do not have an event horizon but a partially reflective membrane is placed slightly outside the usual position of the event horizon for the stability of these compact objects. For a rotating Kerr-like compact object with mass $M$ and angular momentum $J=aM$, the metric in the Boyer-Lindquist coordinate can be written as
\begin{eqnarray} \label{kerr}
  ds^2 =-\big(1-{2Mr \over \rho^2}\big)dt^2+\bigg(r^2+a^2+{2a^2Mr \sin^2\theta \over \rho^2}\bigg)\sin^2\theta d\phi^2
-{4aMr \sin^2\theta \over \rho^2}d\phi dt
 +{\rho^2 \over \Delta}dr^2+\rho^2 d\theta^2,
\end{eqnarray}
where we defined  \begin{align}
~~~~~~\rho^2=r^2+a^2\cos^2\theta, ~~~~~~\Delta=r^2+a^2-2Mr=(r-r_+)(r-r_-).
\end{align}
The position of the classical black hole horizons and the angular velocity of the horizon is given as 
\begin{align}
r_{\pm}=M\pm \sqrt{M^2-a^2},\qquad \Omega_H={a\over 2Mr_+}.
\end{align}In this section, we study the QNM spectrum and the absorption cross-section of the Kerr-like ECO with a modified boundary condition at the horizon. Later we will reproduce these results from a dual CFT analysis by establishing a correspondence between  quantum gravity in the near-horizon region and a thermal two-dimensional CFT.   We study a massless field in the Kerr background and the Klein-Gordon equation for the scalar field is given as:
\begin{align} \label{scalareq}
    \frac{1}{\sqrt{-g}}\partial_{\mu}(\sqrt{-g}g^{\mu \nu }\partial_{\nu}\Phi)=0
\end{align}
We will assume excitation wavelengths larger than the black hole radius  i.e. $\omega M\ll 1$ to simplify the wave equation, which can be solved analytically by using the method of asymptotic solution matching, where the background spacetime is divided into a ``near-region"($\omega r\ll 1$), a ``far-region"($r\gg M$)  and a intermediate matching region($M\ll r\ll 1/\omega$). As shown in Appendix \ref{app_a}, we solve the wave equation in near/far-region and match the solutions in the intermediate region($M\ll r_{match}\ll 1/\omega$) at some arbitrary position( given as ``r$_{match}$"). %

\subsection{Scattering }
\label{flux} 
We use the approximate wave function obtained in Appendix \ref{app_a} to compute the absorption cross-section of a Kerr-like ECO and analyze the effects due to the presence of the near-horizon quantum structure. We assume, at $(r_{h}+\epsilon)$ there is a reflective membrane with reflectivity $\mathcal{R}$, where $\epsilon$ is usually assumed to be of the order of the Planck length. This reflective membrane can be seen as an effective description of the quantum corrections at the horizon such as a firewall or as seen in the fuzzball scenario. The scattering cross-section of the ECO is sensitive to the near-horizon geometry as presence of such reflective membranes would modify the boundary condition at the horizon. 
\begin{figure}[h] 
    \includegraphics[width=11cm]{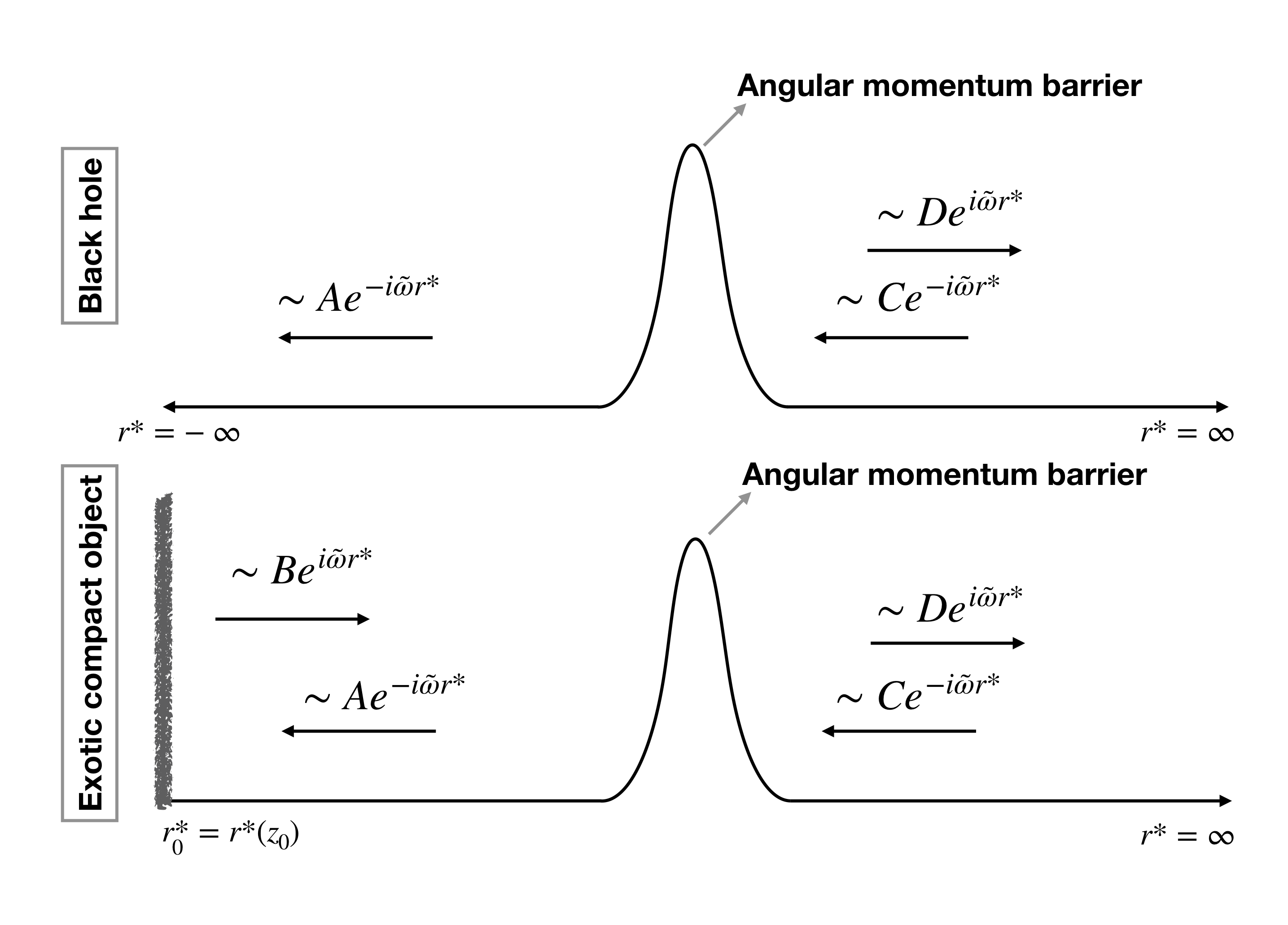}
    \caption[] {{The difference in the near-horizon region between a black hole and ECO is shown in the above schematic diagram. Due to the presence of a quantum structure at $r^*_0$ the near-horizon region would have an additional wave reflected by the membrane placed in front of the would-be horizon. This wave would eventually tunnel through the potential barrier and reach the asymptotic region after a time delay.}}\label{eco}
 \end{figure}
For a scalar perturbation in the Kerr background  the asymptotic behaviour of the waveform would be given as 
\begin{align}\label{asymp_soln}
    R\sim
    \begin{cases} 
      C r^{-1}e^{-i\omega r^*}+Dr^{-1}e^{i\omega r^*}& \text{for}\ \quad r^*\to \infty \\
      Ae^{-i\tilde\omega r^*}+Be^{i\tilde\omega r^*} & \text{for} \quad r^*\to -\infty,
    \end{cases}
\end{align}
where $\tilde \omega$ is the horizon-frame frequency, $r^*$ is the tortoise coordinate defined in \eqref{tor} and the asymptotic amplitudes ($A,B,C,D$) are shown in Fig (\ref{eco}). The reflectivity of the membrane can be defined in terms of these asymptotic amplitudes as 
\begin{align} \label{ref}
\mathcal{R} e^{i\pi\delta}\equiv {B\over A}z_0^{2i\sigma},
\end{align}
where $\delta$ is a phase determined by the quantum properties of the ECO, $\sigma={\omega_H-m\Omega\over4\pi T_H} $  and $z_0$ corresponds to a particular point outside the horizon, where the reflective membrane is placed, in terms of the coordinate defined as $z={r-r_+\over r-r_-}$.  The conserved flux associated to the radial wave equation \eqref{wave_eq_hd}  for the radial wavefunction $R$ is given as 
\begin{align}
F=-i2\pi\big( R^*\Delta\partial_rR-R\Delta \partial_rR^*\big) .
\end{align}
Using this expression we can calculate the ingoing/outgoing flux at the horizon ($F^{in}_{r\to r_+}, F^{out}_{r\to r_+} $) and the ingoing flux from infinity ($F^{in}_\infty$), 
to  obtain the absorption cross-section as 
\begin{align} \label{eco_abs}
\sigma_{abs}={F^{in}_{r\to r_+}+F^{out}_{r\to r_+}\over F^{in}_{r\to \infty}}=4\pi\sigma(r_+-r_-){1-\mathcal{|R|}^2\over |\alpha/A|^2+|\beta/A|^2}.
\end{align}
 To arrive at the last expression we have used the reflectivity of the ECO \eqref{ref} to write the asymptotic amplitude $B$ in terms of $A$ and  $\alpha, \beta$ can be expressed as some linear combination of $C,D$ as given in Appendix \ref{app_a}. As we are performing the calculation in the low frequency limit, $|\beta/A|$ is suppressed compared to $|\alpha/A|$ due to the positive power of $\omega$ in the former. Using $|\alpha/A|$ as given in \eqref{amplitude_rel} we can write the absorption cross-section, \eqref{eco_abs}, as
 
 \begin{align} \label{kerrabs}
&\sigma_{abs}\sim
\nonumber\\
&\omega^{2l+1}\sinh\bigg({4\pi Mr_+\over r_+-r_-}(\omega-m\Omega_H)\bigg)
\left|\Gamma\bigg(1+l-i2M\omega\bigg)\Gamma\bigg(1+l-i{4M^2\over r_+-r_-}\omega+i{2a\over r_+-r_-} m\bigg)\right|^2
{1-|\mathcal{R}|^2\over \left|1-\mathcal{R}e^{-2ir_0^*(\omega_H-m\Omega)+i\delta}\right|^2},
 \end{align}
 where $r_0^*$ is the position of the membrane in terms of tortoise coordinate. As an example, Figure (\ref{fig1}a) shows $\sigma_{abs}$ for an ECO with different constant values of ${\cal R}$ (with ${\cal R} =0$, corresponding to a classical black hole). 
However, like any physical system, the reflectivity can change with frequency, depending on its microscopic structure and energy levels. The Boltzmann frequency-dependent reflectivity for a quantum black hole was first derived in \cite{Oshita:2019sat,Wang:2019rcf} and is given by 
\begin{align} \label{ref_B}
\mathcal{R}\sim e^{-|\omega-m\Omega_H|/2T_{QH}},
\end{align}
where $T_{QH}$ is defined as the ``Quantum Horizon temperature'' \cite{Oshita:2020dox} and as shown, it is expected to be comparable to the Hawking temperature:
\begin{align}
T_H={1\over 8\pi}{r_+-r_-\over Mr_+}.
\end{align}Thus we can write $T_{QH}=\gamma T_H$ , where $\gamma$ is the proportionality constant that depends on the dispersion and dissipation effects in graviton propagation \cite{Oshita:2018fqu}. In order to avoid instability of the ergoregion (for all BH spins) one must satisfy the condition, $\gamma <1.86 $ \cite{Oshita:2020dox}.
In Fig.(\ref{fig1})b we have plotted the absorption cross-section for different positions of the reflective membrane (using the reflectivity defined in Equation \ref{ref_B}) in front of the horizon.

\begin{figure}[h] 
\begin{minipage}{14cm}
  \includegraphics[width=14cm]{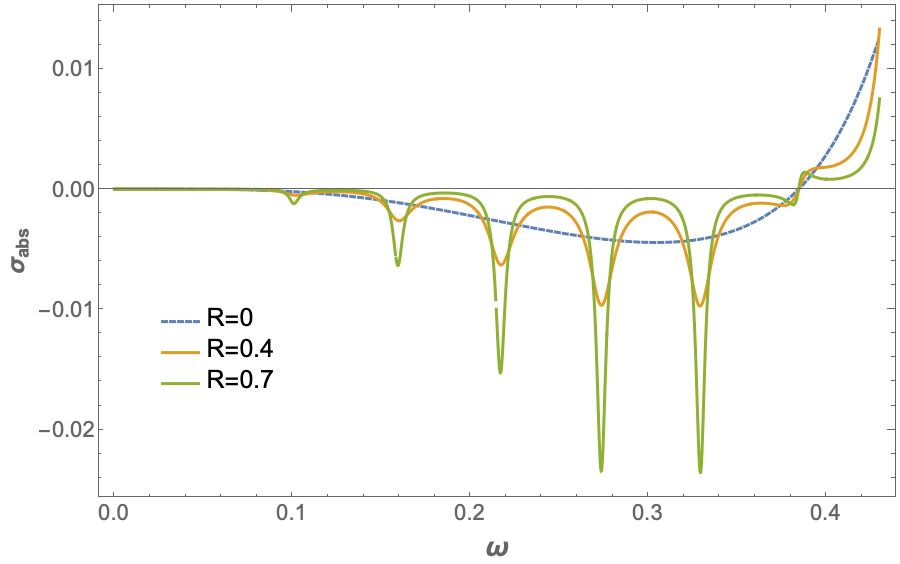} 
   \end{minipage}
   \begin{minipage}{14cm}
   \includegraphics[width=14cm]{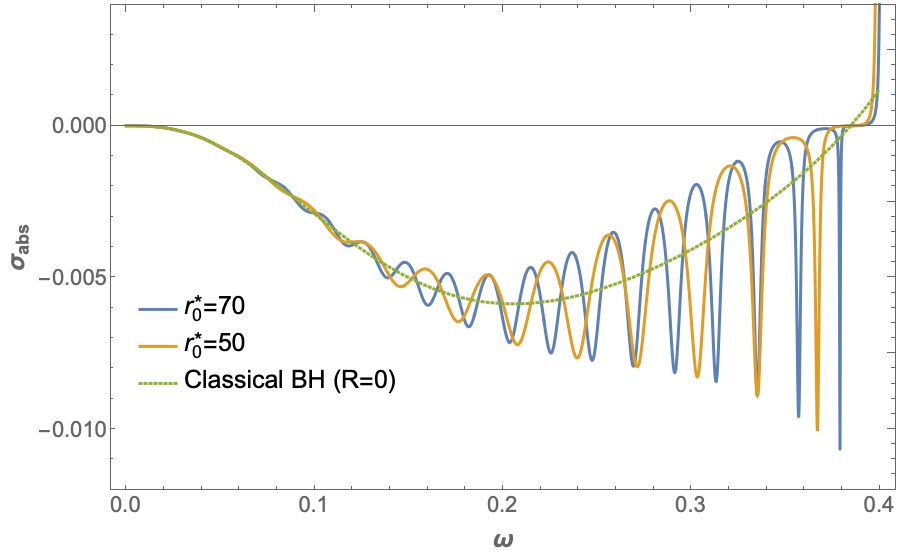} 
   \end{minipage}
 \caption{ The figure on the top shows the absorption cross-section for different values of the reflectivity keeping the position of the membrane fixed at $r_0^*=50$. For the bottom figure we used a frequency dependent reflectivity (Boltzmann reflectivity \cite{Oshita:2019sat}) and plotted the absorption cross-section for different positions of the wall. The dotted line in the bottom figure corresponds to the classical black hole absorption cross-section with zero reflectivity.  For this plot we assumed $\gamma=1$. For both the plots we used $a=0.67, m=2, l=2.$  }\label{fig1}
 \end{figure}

For a Kerr black hole, the low frequency absorption cross-section is negative due to superradiance \cite{Brito:2015oca} $(\omega<m\Omega_H)$ but the novel feature of the absorption cross-section of the ECO's are the oscillatory features, superposed on top of the classical cross-section in Fig. (\ref{fig1}). These oscillations corresponds to resonances at the ECO quasi-normal frequencies.

\subsection{Quasi normal modes}

To obtain the quasi normal modes of a classical black hole, usually an ingoing boundary condition is imposed at the horizon and the boundary condition at asymptotic infinity would be set by the fact that there is no incoming wave from infinity (in Fig\eqref{eco} this would correspond to setting $C=0$). There are various approaches used in the literature to obtain the QNM spectrum of a black hole/ECO. Analytically it is difficult to obtain the QNM of ECOs without making assumptions as the wave equation governing the perturbations are difficult to solve, depending on the model of the ECO. Numerical techniques are used to solve the wave equation or one can obtain the QNM spectrum from the poles of the Green's function expressed in terms of a transfer function\cite{Mark:2017dnq}. 
As demonstrated in Appendix \ref{app_a} we use the low frequency approximation to solve the wave function and obtain the QNM spectrum. 
Using the frequency dependent Boltzmann reflectivity \eqref{ref_B},  in \eqref{qnf}  the QNM spectrum is given as\cite{Wang:2019rcf}
\begin{align} \label{qnf_B}
\omega_{n}-m\Omega_H\sim{\pi (2n+1+\delta)\over 2 r^*_0}\left[1-{i\times \text{sgn}(2n+1+\delta)\over 4r^*_0\gamma T_H}\right].
\end{align}
This matches exactly with the quasi normal frequencies obtained in \cite{Oshita:2019sat}, assuming the Boltzmann reflectivity for the near-horizon membrane in Kerr spacetime. 
One must note that one major difference between the QNM spectra obtained for the ECOs and the black holes is that the quality factor, $Q \equiv \Re\omega_n/\Im\omega_n $ for ECO QNM is parametrically enhanced, and approaches infinity in the continuum limit $r^*_0 T_H \to \infty$.  In contrast, $Q\lesssim 1$ for classical black hole QNMs.

\section{Towards a Holographic Interpretation of Kerr-like ECOs}
\label{hidden} 

\subsection{Hidden conformal symmetry of classical Kerr spacetime}
Let us start by reviewing the current progress in the Kerr/CFT correspondence for non-extremal Kerr black holes. 
Castro, Maloney, and Strominger have found that Kerr black holes with generic mass and spin inherit a hidden local conformal symmetry acting on the low frequency modes \cite{Castro:2010fd}. When solving for the wave equation using the method of the asymptotic solution matching, we saw that the final solution does not depend on $r_{match}$ and this indicates that the solution in the near region must inherit some special symmetry keeping it invariant under transformations of $r_{match}$.
The symmetry becomes evident in the {\it near-horizon} region of the phase space defined as 
$
\omega(r-r_+) \ll 1,
$
where $r$ is the Boyer-Lindquist radial coordinate and $r_+$ is the radius of the outer horizon. The near-horizon wave function for a scalar field in Kerr background is given by hypergeometric functions \eqref{sol_near}, one can understand the emergent conformal symmetry from the fact that these hypergeometric functions fall into the $SL(2,R)$ representation. The conformal symmetry is called hidden in this case as it acts locally on the solution space and it is globally broken to $U(1) \times U(1)$ by the periodic identification of the azimuthal angle $\phi$.

To understand the hidden conformal symmetry explicitly, one can introduce a set of 
conformal coordinates \cite{Castro:2010fd,Haco:2018ske,Chen:2020nyh} (similar to the coordinate transformation that turns Poincare $AdS_3$ into BTZ black hole \cite{KeskiVakkuri:1998nw,Maldacena:1998bw}):   
\begin{align}
 \label{cftcoordinate}
& w^+= \sqrt{r-r_+ \over r-r_-}e^{{2\pi T_R}\phi},
 \\
 &w^-= \sqrt{r-r_+ \over r-r_-}e^{{2\pi T_L}\phi-{t \over 2M}},
 \\
&y= \sqrt{r_+ -r_-\over r-r_-}e^{{\pi(T_R+T_L)}\phi-{t \over 4M}},
\end{align}
where we defined a right and left temperature as
\begin{align}
T_R={r_+-r_-\over4\pi a},~~~~~~T_L={r_++r_-\over4\pi a}.
\end{align}
In terms of the conformal coordinates, the past horizon and the future horizons are at $w^\pm=0$, respectively.
The inverse transformation is given by
\begin{align} \phi&={1 \over 4\pi T_R}\ln{w^+(w^+w^-+y^2) \over w^-},
\\  
r&=r_++4\pi a T_R{w^+w^-\over y^2}, 
\\ 
t&={M(T_R+T_L) \over T_R}\ln{w^+\over w^-}+{M(T_L-T_R)\over T_R}\ln (w^+w^-+y^2).
\end{align}
To leading order around the bifurcation surface, it is possible to obtain a metric in terms of the conformal coordinates from \eqref{kerr} as \cite{Haco:2018ske}
\begin{align}\label{conformalkerr}
ds^2={4\rho_+^2\over y^2}dw_+dw_-+{16J^2sin^2\theta\over y^2 \rho_+^2}dy^2+\rho_+^2d\theta^2+...
\end{align}
Using these coordinate transformations one can express the scalar wave equation in the {\it near-horizon} region as the  $SL(2,R)$ Casimir with conformal weights \cite{Castro:2010fd, Compere:2012jk}
\begin{align}
(h_L,h_R)=(l,l).
\end{align}
As mentioned earlier the $SL(2,R)_L\times SL(2,R)_R$ symmetry is spontaneously broken by the  periodic identification of the angular coordinate
\begin{align}\label{2pi}
\phi \to \phi+2\pi.
\end{align}
For the given periodicity, the conformal coordinates would be identified as   
\begin{align}
\label{idn} w^+ \sim e^{4\pi^2 T_R}w^+, ~~ w^- \sim e^{4\pi^2 T_L}w^-, ~~
y\sim e^{2\pi^2 (T_R+T_L)}y.\end{align}
From \eqref{cftcoordinate}, at fixed radial distance $r$ we can write the relation between the conformal coordinates and Boyer-Lindquist coordinate as
\begin{align} \label{cftrindler}
w^{\pm}=e^{\pm t_{R,L}}.
\end{align} 
This relation looks analogous to the relation that identifies the Minkowski coordinates $(w^{\pm})$ to the Rindler coordinates$(t_{R,L})$ where we defined 

\begin{align} \label{monocoordinate}
&t_R=2\pi T_R\phi,
\nonumber\\
&t_L={t\over 2M}-2\pi T_L\phi.
\end{align}
The periodic identification of the angular coordinate $\phi$ in \eqref{monocoordinate} gives the periodicity of these coordinates as
 \begin{align} \label{cor_periodicity}
 (t_L,t_R)\sim (t_L,t_R)+4 \pi^2(-T_L,T_R).
 \end{align}
 The frequencies $(\omega_L,\omega_R)$ associated with the Killing vectors $(i\partial_{t_L},i\partial _{t_R})$ are conjugate to $(t_L,t_R)$. We can write the relation between $(\omega,m)$, which are eigenvalues of the operators $(i\partial _t,-i\partial _{\phi})$, and $(\omega_L,\omega_R)$ through \cite{Castro:2013kea}
 \begin{align}
 e^{-i \omega t+im\phi}=e^{-i \omega_Lt_L-i\omega_Rt_R}.
 \end{align}
 This relation between the frequencies along with \eqref{monocoordinate} would explicitly give the left/right frequencies as
 \begin{align} \label{qnf_matching0}
&\omega_L=2M\omega,
 \\
 & \omega_R={2M^2\over \sqrt{M^2-a^2}}\omega-{a\over \sqrt{M^2-a^2}}m.
 \end{align}
As given in \cite{Castro:2010fd}, another way of writing the relation between the left, right frequencies ($\omega_L,\omega_R$) and $(\omega,m)$ of Kerr spacetime having an entropy $S_{BH}$, Hawking temperature $T_H $, is by taking a thermodynamic route using the first law of thermodynamics, $T_H\delta S_{BH}=\delta M-\Omega \delta J $. Identifying $\omega=\delta M$ and $m=\delta J$, we need to consider the conjugate charges ($\delta E_R, \delta E_L$), following the relation
\begin{align}
\delta S_{BH}={\delta E_R \over T_R}+{\delta E_L \over T_L},
\end{align}to show the relation between $(\omega,m)$ and $(\delta E_R, \delta E_L)$ as\footnote{One must note that $\delta E_{L,R}=\tilde \omega_{L,R}$ is  conjugate to the re-scaled coordinate $\tilde t_{L,R}=t_{L,R}/2\pi T_{L,R}$ (in terms of which we will later define the torus coordinates \eqref{torus_coordinate} on which the CFT lives), thus drawing the equivalence between this identification and \eqref{qnf_matching0}.}
\begin{align} \label{qnf_matching}
&\delta E_L=\tilde\omega_L= \omega_L 2\pi T_L ={2M^3\over J}\omega,
\nonumber
\\
&\delta E_R=\tilde\omega_R= \omega_R 2\pi T_R ={2M^3\over J}\omega-m.
\end{align}
In the context of Kerr/CFT correspondence this identification between the frequencies were used to show the consistency between the absorption cross-section of the  dual CFT (calculated from the thermal two-point function) and the calculation performed in Kerr spacetime \cite{Castro:2010fd,Chen:2010xu}. 

\subsection{Dual CFT of the Kerr-like ECO}

Now that we have reviewed the current understanding of the classical Kerr/CFT correspondence, we shall derive a CFT dual to a Kerr-like ECO. 
As we have discussed in Section \ref{ECO},
for a Kerr-like ECO the horizon is modified due to  quantum gravitational effects in the near-horizon region of a black hole, while the exterior spacetime is well approximated by the Kerr metric.  The equations governing the evolution of the perturbation in the ECO background is exactly the same as in classical Kerr spacetime; the only difference is in the boundary condition at the horizon. Hence, in the case of the ECO, the solution of the near-horizon scalar perturbation is given in terms of hypergeometric functions as well (hinting at the hidden conformal symmetry). One can define a set of vector fields
 (similar to the classical Kerr case \cite{Castro:2010fd}) falling in the $SL(2,R)$ representation and the Casimir would be the same as the near-horizon scalar perturbation equation, allowing us to identify the conformal dimension of the CFT. 
Usually, quantum gravity modifications to the bulk geometry can be understood as  non-perturbative effects in the dual field theory,  finite $N$ and/or finite size effects in the usual AdS/CFT terminology \cite{Kabat:2014kfa,Birmingham:2002ph,Solodukhin:2005qy}. We can then use the fact that the dual CFT lives in a space of finite volume to reproduce the discrete QNM spectrum of the Kerr-like ECO. We can see similar finite $N$/finite size effects in the holographic interpretation of the brick wall \cite{Iizuka:2013kma,Solodukhin:2005qy}; the near-horizon cut-off can be understood in terms of finite $N$ effects in the dual CFT  in order to make the free energy and the entropy finite (on either side of the duality).

In order to compute the QNM spectrum and the absorption cross-section from the dual CFT, we start with the thermal two-point function of a CFT living on an Euclidean torus having a spatial cycle of length $L$ and an temporal cycle of length $1/T$. Unlike in the limit of, $L\gg 1/T$ or $L\ll 1/T$ where the torus decompactifies to a cylinder, it is usually difficult to determine the two-point function on a torus keeping both the lengths finite, as conformal symmetry is not enough to determine the universal form of the correlator.   As shown by \cite{Maldacena:2001kr,Kleban:2004rx,Birmingham:2002ph,Solodukhin:2005qy}, some progress can be made in special cases such as the supersymmetric CFT in the strong coupling regime dual to string theory in $AdS_3$ which describes the low energy excitation of D1-D5 branes \cite{Aharony:1999ti}.

 In the usual description of Kerr/CFT correspondence, the cylinder approximation of the torus is considered based on the assumption $L\gg 1/T$  and thus the spatial direction effectively decompactifies \cite{Castro:2010fd,Maldacena:1997ih}. In the context of $AdS_3/CFT_2$, we have seen that a similar approximation is done for the BTZ black holes whereas, for its T-dual, thermal $AdS_3$ one can assume $1/T\gg L$. However, to include the finite size/$N$ effects, we can no longer use the $L\gg1/T$ approximation. Therefore, for computation of physical quantities such as the absorption cross-section,  the precise doubly-periodic two-point function on the torus is needed. The latter is computed in Appendix \ref{2point}.

One way to put the dual CFT on a circle of finite length, in the context of Kerr CFT,  is  by considering a rescaling of the temperatures$(T_R,T_L)$, that appear in the conformal coordinates \eqref{cftcoordinate}, by some constant. 
One can verify the consistency of the Cardy entropy on rescaling of the temperatures in the conformal coordinates from the recent arguments given in \cite{Chen:2020nyh}. We modify the size of the covering space, where the CFT lives, by introducing a factor of $L$ ($L$ is an integer) in the periodicity of the azimuthal angle that appears in the torus coordinates. As we will see, $L$ will correspond to the length of the circle on which the CFT lives, thus allowing us to study the finite size/N effects. We must warn the reader that the gravity/CFT duality is not well defined in the finite size/N limit and hence, to a certain extent, we have to guess  how the finite size/N effects of the holographic CFT reflects on the gravity side. Our objective is to verify if the QNM spectrum (calculated from the poles of  two-point function of a CFT living on a torus)  matches with the spectrum obtained on the gravity side, once the black hole horizon is removed and replaced by a partially reflective membrane in front of the would-be horizon.

For the ECO the usual periodic identification of the azimuthal coordinate is there, as given in \eqref{2pi}. We can combine this azimuthal periodicity (times an integer multiple of $L$) with thermal periodicity of imaginary time to get
 \begin{align}\label{eco_period}
&\phi \to \phi + 2L \pi +i\Omega_H/T_H,
\\
&t=t+i/T_H.
\end{align}
Now, under the periodic identifications \eqref{eco_period} for the coordinates of the torus \eqref{monocoordinate} we get
 \begin{align}\label{cor_periodicity_new}
 ( t_L, t_R)\sim ( t_L, t_R)+(-4 \pi^2T_LL-i(2\pi {T}_L\Omega_H/T_H-1/2MT_H),4\pi^2{T}_RL+i2\pi {T}_R\Omega_H/T_H).
 \end{align}

\subsection{Quasi normal modes spectrum from the dual CFT}

The postmerger ringdown phase after a binary black hole coalescence is described in terms of QNM and can carry information about the near-horizon quantum structures of an ECO. In the context of AdS/CFT, it was shown in \cite{Birmingham:2001pj} that the black hole QNM spectrum can be obtained from the poles of the retarded CFT correlation function. Assuming that quantum gravity in the near-horizon region of a Kerr black hole  is dual to a two-dimensional thermal CFT, as conjectured by the Kerr/CFT correspondence, the consistency between the gravity results and CFT computations of the QNM was shown in \cite{Chen:2010xu,Chen:2010ni}.

  Here, we conjecture that quantum gravity living in the near-horizon region of a Kerr-like ECO is dual to a two-dimensional CFT that lives on a circle of finite length $L$, where the CFT coordinates are defined by the relation \eqref{monocoordinate} having periodicities as given in \eqref{cor_periodicity_new}. To obtain the QNM spectrum from the poles of the retarded  CFT correlator, we need to perform the Fourier transform of the retarded correlation function to momentum space.  It is complicated to perform this computation due to the presence of the $\Theta-$function in the retarded two-point correlation function. Instead, we perform the Fourier transform of the two-point function given in
  \eqref{2pointtorus} and look at the poles lying in the lower half-plane as these would match with the poles of the retarded correlation function. The Fourier transform of the two-point function is given as \footnote{ The infinite sum appears  in front of the two-point function as we are implementing the method of images. To describe an ECO/black hole in the dual CFT one must take into account the correct periodic identification as given in \eqref{cor_periodicity_new} and hence we are shifting the periodicity by an integer multiples and summing over the images  }
\begin{align}
&\bar G(\omega_L,\omega_R)=\int d t_R \,d t_L\,e^{-i\omega_R  t_R}e^{-i\omega_L  t_L}\langle\mathcal{O}( t_R, t_L)\mathcal{O}(0,0)\rangle_{torus} 
\nonumber\\
&=\int d t_R \,d t_L \,e^{-i\omega_R  t_R}e^{-i\omega_L  t_L} \sum_{p\in \mathbb{Z}}{(\pi T_R)^{2h_R}(\pi T_L)^{2h_L} \over[\sinh({ t_R\over 2}+p(2\pi ^2LT_R+i\pi T_R{\Omega_H\over T_H}))]^{2h_R}[\sinh({ t_L\over 2}+p(2\pi^2 L T_L-i\pi T_L{\Omega_H\over T_H}+{i\over 4MT_H}))]^{2h_L}}
\nonumber\\
&=\int d t_R \,d t_L \,e^{-i\omega_R  t_R}e^{-i\omega_L t_L} \sum_{p\in \mathbb{Z}}{(\pi T_R)^{2h_R}(\pi T_L)^{2h_L} \over[\sinh[\pi T_R({ t_R\over 2\pi T_R}+p(2 \pi L +i/T_R))]]^{2h_R}[\sinh[\pi T_L({ t_L\over 2\pi T_L}+p(2 \pi L+i/T_L ))]^{2h_L}}. \label{fourier2pointtorus2}
\end{align}
The CFT two-point function derived here based on the torus coordinate along with their periodicities \eqref{cor_periodicity_new} matches with  the generic two-point function of a CFT living on a Euclidean torus as given in \eqref{2pointtorus} if we  choose the group parameters as $(\mathbf{a}=1,\mathbf{b}=1,\mathbf{c}=-1,\mathbf{d}=0)$.  However,the CFT two-point function can be written in a more general way by keeping the value of the group parameter $\mathbf{a}$ arbitrary.  
Without fixing the value of $\mathbf{a}$, the two-point  function \eqref{fourier2pointtorus2} is now modified as 

\begin{align} \label{fourier2pointtorusA}
&\bar G(\omega_L,\omega_R)=
\nonumber\\
&\int d t_R \,d t_L \,e^{-i\omega_R  t_R}e^{-i\omega_L t_L} \sum_{p\in \mathbb{Z}}{(\pi T_R)^{2h_R}(\pi T_L)^{2h_L} \over[\sinh[\pi T_R({ t_R\over 2\pi T_R}+p(2 \pi L +i\mathbf{a}/T_R))]]^{2h_R}[\sinh[\pi T_L({ t_L\over 2\pi T_L}+p(2 \pi L+i\mathbf{a}/T_L ))]^{2h_L}}.
\end{align}
In order to do the Fourier transform in a convenient way we perform a coordinate transformation, defining the torus coordinates $(\tilde t_{R,L})$ as
\begin{align}\label{torus_coordinate}
&\tilde t_{R}= t_{R}/2\pi T_R+p(2\pi L+i\mathbf{a} /T_R).
\nonumber\\
&\tilde t_{L}= t_{L}/2\pi T_L+p(2\pi L  +i\mathbf{a} /T_L).
\end{align}
 These newly defined coordinates along with \eqref{fourier2pointtorusA} would give us

\begin{align} \label{2pointtorusk}
&\bar G(\tilde \omega_L,\tilde \omega_R)\sim
\nonumber\\
&\sum_{p\in \mathbb{Z}}
e^{ip (2 \pi \tilde\omega_LL+2 \pi \tilde\omega_RL+i\mathbf{a} {\tilde\omega_R\over T_R}+i\mathbf{a}{\tilde\omega_L\over T_L})}
\int d\tilde t_{R} \,d \tilde t_{L} \,e^{-i\tilde\omega_R  \tilde t_{R}}e^{-i\tilde\omega_L \tilde t_{L}} {(\pi T_R)^{2h_R}(\pi T_L)^{2h_L} \over[\sinh(\pi T_R\tilde t_{R})]^{2h_R}[\sinh(\pi T_L\tilde t_{L})]^{2h_L}}
\nonumber\\
&\propto 
T_{L}^{2h_L-1}T_{R}^{2h_R-1}e^{-{\tilde\omega_L\over 2T_L}-{\tilde\omega_R\over 2T_R}}\left|\Gamma\bigg(h_R+i{\tilde\omega_R\over2\pi T_R}\bigg)\Gamma\bigg(h_L+i{\tilde\omega_L\over2\pi T_L}\bigg)\right|^2\nonumber\\&
\hspace{5cm}\times\bigg[
{1\over 1-e^{i2\pi L(\tilde\omega_R+\tilde\omega_L)-\mathbf{a}\left|{\tilde \omega_R\over T_R}+{\tilde \omega_L\over T_L}\right|}}
-{1 \over 1-e^{i2\pi L(\tilde\omega_R+\tilde\omega_L)+\mathbf{a}\left|{\tilde \omega_R\over T_R}+{\tilde \omega_L\over T_L}\right|}}\bigg]
.
\end{align}
The CFT two-point function in the  momentum space has two set of  poles coming from the exponential part of \eqref{2pointtorusk} lying in the upper and lower half of the $\omega$ plane. As we discussed above, the QNM spectrum is given by the poles of the retarded correlation function, the poles of \eqref{2pointtorusk} in the lower half plane are the ones relevant for us and these are given as\footnote{There are other poles coming from the singularities of the Gamma function and it can be shown that these poles match with the usual Kerr black hole QNM spectrum obtained by imposing an ingoing boundary condition at the horizon.}
\begin{align}\label{torusqnf}
2\pi L(\tilde \omega_L+\tilde\omega_R)+i\mathbf{a}\bigg|{\tilde \omega_R\over T_R}+{\tilde \omega_L\over T_L}\bigg|=2\pi n.
\end{align}
Using the relation between the CFT frequencies and the ECO frequencies as given in \eqref{qnf_matching} we get
\begin{align} \label{QNM_CFT}
&{4M^2\over a}\omega-m+i\mathbf{a}\bigg({1\over 2\pi L T_H}|\omega-m\Omega_H|\bigg)={\pi\over 2\pi L}2n,
\nonumber\\
& \implies \omega-m\Omega_H\simeq{a\over  8M^2 L}(2n+1+\delta)\bigg[1-i\mathbf{a}{a\times ~\text{sgn}[2n+1+\delta]\over 8M^2\pi LT_H }\bigg].
\end{align}
where we have defined the quantity $\delta$ as
\begin{align}\label{phase}
\delta=-2mLr_-/ r_+-1.
\end{align}
For the Kerr-like ECO the QNM spectrum is calculated in the low frequency approximation as given in \eqref{qnf_B} considering a Boltzmann reflectivity of the near-horizon membrane   \cite{Wang:2019rcf}. 
We notice that {\it both} the real and imaginary parts of the quasi normal frequency for Boltzmann ECOs \eqref{qnf_B}   would match with the CFT result if we take, $\mathbf{a}=1/2\gamma$ and the length of the torus as
\begin{align} \label{echotimeL}
L={a|r^*_0|\over  4\pi M^2}.
\end{align}
Comparing \eqref{QNM_CFT} with \eqref{qnf} we can explicitly read off the reflectivity of the membrane as
\begin{align}
    \mathcal{R}= e^{-|\omega-m\Omega_H|\mathbf{a}/T_H},
\end{align}
which is identical to the generalized Boltzmann reflectivity introduced in \cite{Oshita:2020dox}, reducing to the standard Boltzmann  for ${\bf a} =1/2$ \cite{Oshita:2019sat}.
Furthermore, from \eqref{phase} we get the phase of reflection as defined in \eqref{ref}. The phase can depend on the frequency  but as we are working in the low frequency limit such frequency dependence is not evident in our calculation and we determine $\delta$ as a constant.

\subsection{CFT interpretation of Absorption probability}

In the previous section, we saw that the identification \eqref{echotimeL} makes the QNM of the ECO consistent with the dual CFT computation.  The other important check about the accuracy of this identification is to compute the absorption cross-section from the CFT two-point function and see if it can determine the near-horizon contribution to the scattering cross-section correctly, as predicted by the gravity result in \eqref{kerrabs} (Fig (\ref{fig1})). The low frequency absorption cross-section for a massless scalar in Kerr spacetime is a well known result \cite{Maldacena:1997ih}.  In \cite{Castro:2010fd,Maldacena:1997ih,Bredberg:2009pv} the absorption cross-section was computed for a scalar field in Kerr background with an ingoing boundary condition at the horizon and its equivalence with the CFT computation was shown. 

In this section we will compute the absorption cross-section from the CFT two-point function given in \eqref{2pointtorus}. We again consider the dual field theory  on a circle of length $L$ and study the finite size effects of the boundary theory on the absorption cross-section.
 Given the two-point function of a two-dimensional CFT, $G( \tilde t_R, \tilde t_L)=\langle\mathcal{O}( \tilde t_R,  \tilde t_L)\mathcal{O}(0,0)\rangle$, having right and left moving coordinates $( \tilde t_R, \tilde t_L)$, the absorption cross-section can be defined using Fermi's golden rule (to leading order in perturbation theory) as
\begin{align} \label{cft_abs}
\sigma_{abs} \sim 
\int d t_R ~d t_L e^{-i \omega_R  t_R-i\omega_L t_L}[G( t_R-i\epsilon, t_L-i\epsilon)-G( t_R+i\epsilon, t_L+i\epsilon)].
\end{align}
We perform the integral in the same way as we obtained the momentum space two-point function in \eqref{2pointtorusk}. In this case the $\pm i\epsilon$ prescription determined which poles contribute while performing the integral.
 The absorption cross-section is given as:\footnote{Following \cite{Cvetic:2009jn} we can fix the prefactor on the right side of \eqref{momabs}  based on the fact that the emission of any quanta in the dual CFT takes place due to an interaction, between the bulk modes and the CFT operator $\mathcal{O}^{(h_R,h_L)}$, given as
$
\phi_{bulk}\mathcal{O}^{(h_R,h_L)}.
$ It was argued that due to the presence of derivatives acting on the outgoing wave function in the general form of the coupling and the normalization of the outgoing modes, the prefactor would be $(\omega)^{2l-1}$.}
\begin{align} \label{momabs}
&\sigma_{abs} \sim 
\nonumber\\
&T_L^{2h_L-1} T_R^{2h_R-1}(\omega)^{2l-1}\text{sinh}\bigg({\tilde\omega_R \over 2T_R} +{\tilde\omega_L \over 2T_L}\bigg)
\left|\Gamma\bigg(h_R+i{\tilde\omega_R\over2\pi T_R}\bigg)\right|^2\left|\Gamma\bigg(h_L+i{\tilde\omega_L\over2\pi T_L}\bigg)\right|^2
{1-e^{-2\mathbf{a}\left|{\tilde\omega_R \over T_R} +{\tilde\omega_L \over T_L}\right|}\over \left|1-e^{i2\pi L(\tilde\omega_R+\tilde\omega_L)-\mathbf{a}\left|{\tilde \omega_R\over T_R}+{\tilde \omega_L\over T_L}\right|}\right|^2}
\end{align}
In section \ref{flux}, we obtained the low frequency scattering cross-section  by using a matching procedure where we divided the Kerr spacetime into a near-horizon region $(\omega r\ll 1)$, far region $(r\gg M)$ and an overlapping matching region $(1/\omega\gg r\gg M)$. Using the relation between the CFT frequencies $(\tilde \omega_R,\tilde \omega_L)$ and $(\omega,m)$, as given in \eqref{qnf_matching}, along with \eqref{echotimeL}  and \eqref{phase} we get the absorption cross-section as given in \eqref{kerrabs}.  Plotting \eqref{momabs} for the value of  $L$ given by the condition \eqref{echotimeL}, corresponding to a particular position of the membrane, we can see the  correspondence with the gravity result in a better way. 

\begin{figure}[h] 
\includegraphics[width=15cm]{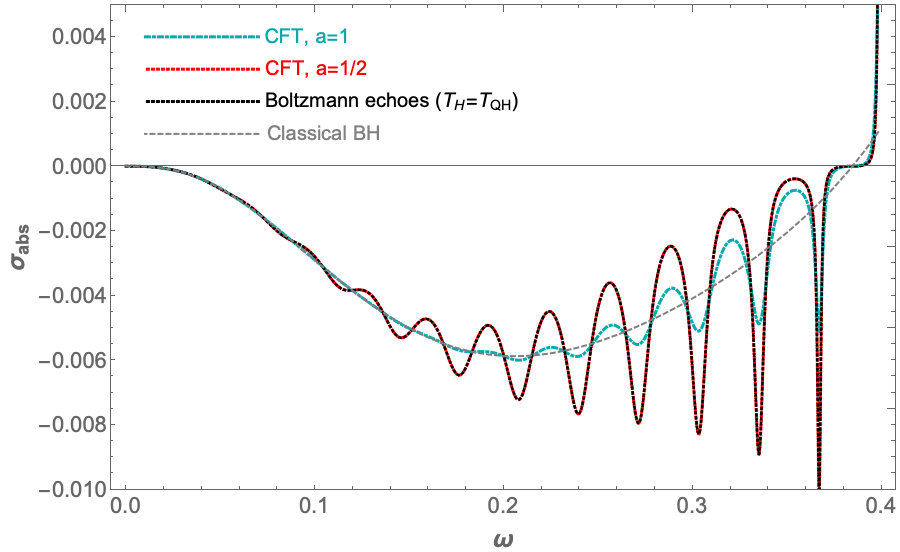}
 \caption[] {{Comparison of the CFT absorption cross-section corresponding to $r^*_0=50$ for the position of the reflective membrane in front of the ECO. For the gravity computation we choose $\gamma=1$ so that $T_{QH}\sim T_{H}$. For the CFT cross-section we see that $\mathbf{a}=1/2$ exactly matches with the gravity computation as we expected. For $\mathbf{a}=1$ we can infer that the reflectivity is suppressed. Once again we assumed $a=0.67, m=2, l=2$ for the plots} }\label{figure:abs_cft}
 \end{figure}

\begin{figure}[h] 
\includegraphics[width=15cm]{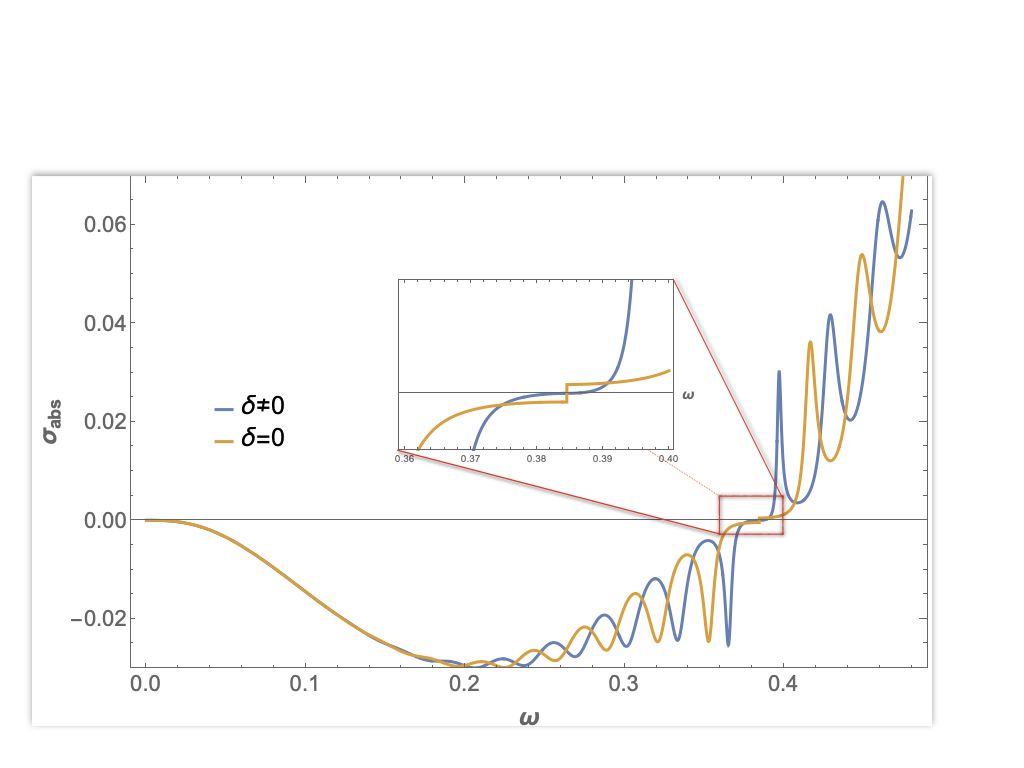}
 \caption[] {{The absorption cross-section for two different values of the phase ($\delta$), placing the reflective membrane at $r_0^*$=50, is shown. The magnified part shows the behaviour of the absorption cross-section at the superradiant frequency, $\omega=m\Omega_H$. }}

\label{phase_plot}
\end{figure}

 In \eqref{phase} we found the phase change due to the reflection at the near-horizon membrane from the dual CFT analysis. 
In Fig (\ref{phase_plot}) we have plotted the absorption cross-section taking $\delta=0$ and the specific value that we found in \eqref{phase}. It is interesting to note that the behavior of the absorption cross-section is different at the superradiant frequency ($\omega=m\Omega_H$) for the two values of $\delta$ that we have chosen for the plot. For $\delta=0$ there is a discontinuity at the superradiant frequency as the Boltzmann reflectivity becomes unity and is non-analytic at $\omega=m\Omega_H$. On the contrary when $\delta=-2mLr_-/ r_+-1$  the discontinuity disappears.

\section{Observable signatures }
\label{observation}
It is well-established  that possible near-horizon modifications to the Kerr spacetime could have signatures in terms of echoes in the ringdown phase of the detected gravitational wave signal \cite{Cardoso:2016rao,Oshita:2018fqu,Abedi:2020ujo}. In the presence of the partially reflective membrane in front of the horizon, we saw that the ringdown of the black hole is modified, leading to a new tower of QNMs for the ECO \eqref{qnf_B}. These QNMs will manifest as repeating echoes, with amplitudes decaying as a power-law \cite{Wang_2018,Wang:2019rcf}, in contrast to the fast exponential decay in case of the classical black hole.  The time delay between two consecutive echoes would give  the echo time-delay and can be written in terms of, $r^*_0$, the distance between the photon sphere and the near-horizon membrane in tortoise coordinates:
\begin{align}\label{techo}
\Delta t_{echo}=2|r^*_0|.
\end{align}
In geometric optics approximations, one can understand this echo time-delay as the time it would take for the classical QNM, generated at the photon sphere due to the perturbation of the ECO, to travel to the membrane placed in front of the horizon and then reflected back to the photon sphere. This argument, as well as the expression \eqref{techo} makes it evident that $\Delta t_{echo}$ is sensitive to the position of the membrane. Converting the tortoise coordinate to the proper distance of the membrane from the would-be horizon $d_{wall}$, Eq. (\ref{techo}) becomes:
\begin{align}\label{echotime}
\Delta t_{echo}\simeq {4 Mr_+\over  r_+-r_-}\ln\left[{8M^2r_+\over (r_+-r_-)d_{wall}^2}\right]={1\over 2\pi T_H}\ln\left[{M\over \pi T_H d_{wall}^2}\right].
\end{align}
In terms of the echo time-delay \eqref{echotime} we can write the length of the circle as
\begin{align}
L=\Delta t_{echo}{a\over 8\pi M^2}=
{a\over 16\pi^2 M^2 T_H}\ln\left[{M\over \pi T_H d_{wall}^2}\right] \in \mathbb{N}. 
\end{align}
From this relation the position of the wall ($d_{wall}$) can be written as 
\begin{align} \label{dwall}
    d_{wall}=\sqrt{{M\over \pi T_H}}e^{-L16\pi^2M^2T_H/a}.
\end{align}
Further, $a,M$ can be written in terms of $S_{BH},T_H$ as
\begin{align}
    &M=T_HS_{BH}\left[-1+\left(1+{1\over 2\pi T_H^2 S_{BH}}\right)^{1/2}\right],
    \\
    &a=T_H S_{BH}\left[\left(-1+\left(1+{1\over 2\pi T_H^2 S_{BH}}\right)^{1/2}\right)^2-4\right]^{1/2}.
\end{align}

\begin{figure}[h] 
\includegraphics[width=14cm]{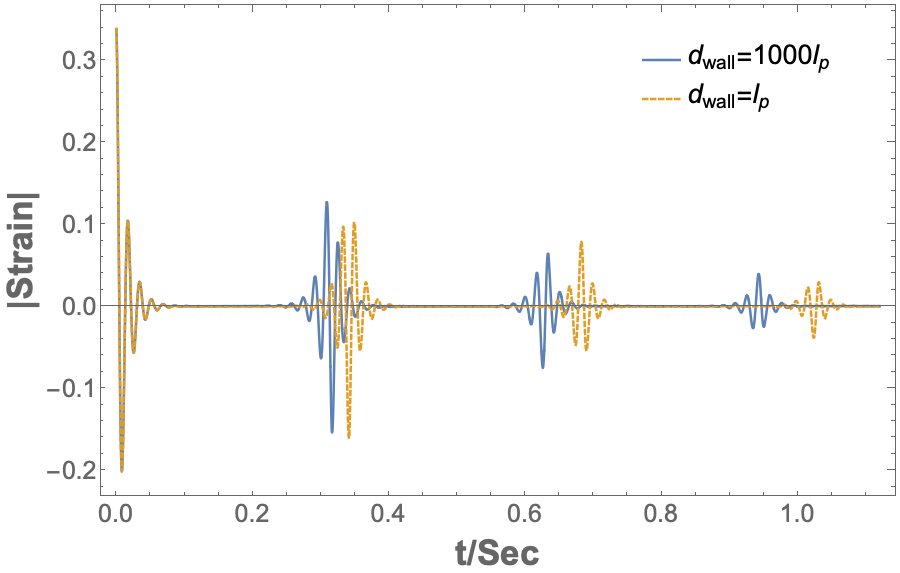}
 \caption[] {{ Echoes in the ringdown for different positions of the reflective membrane placed in front of the would-be horizon. It is often assumed that the membrane is few Planck lengths, $l_p$, away from the horizon. }}

\label{echoplot}
\end{figure}

From \eqref{dwall} we can see that the position of the membrane ($d_{wall}$) in front of the horizon is determined by the size of the covering space $L$, on which the dual field theory lives. If the membrane is pushed all the way to the horizon, i.e. $d_{wall}\to 0$, we get $L\to \infty$, which is the well-defined classical limit on both sides of the duality. Also we note that $L$ is an integer and this puts a constraint on the relation between the echo time-delay and BH properties. In Fig (\ref{echoplot}) we show how the ringdown of the ECO is sensitive to the value of $d_{wall}$. We see that for a larger $d_{wall}$, the distance between the angular momentum barrier and the partially reflective membrane in front of the would-be horizon decreases, leading to shorter echo time-delays.

\section{Discussion}
\label{conclusion}

In this paper, we analysed the holographic CFT dual to a Kerr spacetime whose near-horizon region is modified  due to quantum gravitational effects. One motivation for such near-horizon modification is to address the information loss paradox and we have seen models (such as the firewall or fuzzball) where the horizon is removed or replaced by a partially reflective membrane a few Planckian distance away from the position of the would-be horizon. According to holographic renormalization group (RG) flow program, any UV (near-horizon) modifications to the bulk geometry gets mapped onto an IR modification in the dual CFT \cite{Ishikawa:2013wf,Faulkner:2010jy,Cai:2007zw}. To be more precise, for the Kerr-like ECO, removal of the horizon can be interpreted as a UV cutoff on the gravity side. In this context, we proposed that the dual field theory  lives on a finite \emph{toroidal} two-manifold, where  the limit $LT \to \infty$ for the periodicity of the torus in the spatial and temporal direction is not taken (i.e the spatial circle has a finite length given as $L$). In this picture, the spatial length of the torus, $L$, acts as an IR regulator on the field theory side, and as we do not know the microscopic realization of the Kerr/CFT correspondence, we kept the arguments simply in terms of the basic features of a thermal CFT. One might interpret this as a non-perturbative or finite $N$ effect in the dual field theory which  captures the near-horizon modifications of the Kerr black hole as expected. 

We showed the consistency between the quasi normal modes and the absorption cross-section computed from the gravity side and  similar quantities computed from the dual CFT. On the gravity side, this leads to echoes in the ringdown which repeat with a period of $\Delta t_{echo}$ (which in turn depends on the position of the reflective membrane), and decay as a power law. We  proposed a relation between the length of the circle, $L \in \mathbb{N}$ on which the CFT lives and the echo time-delay,  so that there is an exact matching of the CFT results with the gravity computation. 
This relation between the echo time-delay and $L$ determines the position of the membrane in front of the horizon in terms of the length of the covering space on which the CFT lives. 

For an ECO, the reflectively of the membrane depends on the quantum properties of the near-horizon membrane. Since a rigorous derivation of the reflectivity from a quantum theory of gravity is not available one needs to make a guess about this feature of the membrane. The reflectivity is usually assumed to be a constant, or as proposed recently based on a semi-classical analysis,  the reflectivity is given in terms of the frequency dependent Boltzmann factor \cite{Oshita:2019sat}. From the dual CFT computation, we could determine what the reflectivity would be and surprisingly it does take the form  of the Boltzmann factor. We believe the derivation of the reflectivity of the membrane based on the holographic CFT opens up the possibility of its interpretation in terms the quantum theory of gravity in the near-horizon region of a black hole/ECO. 
The ingoing modes would also undergo a phase change upon reflection by the near-horizon membrane. This phase depends on the intrinsic quantum properties of the membrane, which also follows from our CFT computation. Furthermore, we showed the absorption cross-section has a discontinuity at the superradiance frequency when the phase change at the near-horizon membrane is taken to be zero. However, if we use the the phase predicted by the CFT computation, the absorption cross-section is well-behaved at superradiant frequency.

The Kerr/CFT correspondence was originally established for the NHEK geometry in \cite{Guica:2008mu} and later a microscopic description of the Kerr/CFT for the extremal case was established in \cite{Guica:2010ej,Compere:2010uk} by embedding the NHEK geometry in string theory. For the near-horizon quantum gravity modifications to the Kerr geometry, we saw that in the corresponding dual CFT  one needs to incorporate finite size effects and we can conjecture that such effects would come from finite N corrections in the usual CFT terminology. Understanding the quantum corrections to the ``near-NHEK'' geometry from the CFT side (e.g., \cite{Bredberg:2009pv}) as a result of which the horizon may be removed (e.g., \cite{Solodukhin:2005qy,Bueno:2017hyj}),  would make it possible for a more rigorous analysis in terms of the microscopic theory.

\section*{Acknowledgment }
We would like to thank Steve Carlip, Alejandra Castro, Geoffrey Compère, Lin-Qing Chen, Wan Zhen Chua, Malcolm Perry,  Andy Strominger and Stefano Liberati for invaluable discussions. 
Research at Perimeter Institute is supported in part by the Government of Canada through the Department of Innovation, Science and Economic Development Canada and by the Province of Ontario through the Ministry of Colleges and Universities.

\appendix
\section{Scalar wave equation in Kerr background \label{app_a}}
In this appendix we solve the scalar perturbation equation  using the method of asymptotic solution matching.   The spacetime outside a rotating black hole/ECO is divided into a near-horizon region $(\omega r\ll 1)$, the asymptotic far-region $(r\gg M)$ and a matching region $(1/\omega\gg r\gg M)$ where the near and far-region overlaps. The solutions are obtained in the low frequency limit $1/\omega\gg M$ as the Compton wavelength of the scalar particle is much larger than the typical size of the ECO. We impose suitable boundary conditions at asymptotic infinity and at the horizon to obtain the QNM spectrum and the absorption cross-section of a Kerr-like ECO.
For a massless scalar field satisfying \eqref{scalareq}, with the ansatz 
\begin{align}
\Phi=e^{i(m\phi-\omega t)}S^A(\theta,a\omega)R(r),
\end{align} the radial part of the scalar perturbation equation is given as
\begin{align}\label{wave_eq_hd}
\Delta\partial_r\big(\Delta \partial_r R\big)+ [\omega(r^2+a^2)-ma]^2R-\Delta\lambda R=0~. 
\end{align}
where $\lambda=l(l+1)+a^2\omega^2-2m\omega a$. 
\subsection{Far-region solution}
In the far-region $(r\gg M)$ the scalar perturbation equation reduces to a wave equation for a scalar field having frequency $\omega$ and angular momentum $l$ in flat background,
\begin{align}\label{wave_eq_infty}
\partial_r\big(r^2 \partial_r R\big)+\big[r^2 \omega^2-l(l+1)\big]R=0~.
\end{align}This  equation can be solved in terms of spherical Bessel functions as,

\begin{align}
\label{infty_soln}
R=\frac{1}{\sqrt{r}}\left\{\alpha J_{l+1/2}(\omega r)+\beta J_{-l-1/2}(\omega r) \right\}~,
\end{align} 
where $\alpha$ and $\beta$ are arbitrary constants. In the intermediate matching region  the small $r$ limit of the above equation is relevant and it is given as
\begin{align}
 \label{soln_infty_small}
R\sim\alpha\frac{(\omega/2)^{l+1/2}}{\Gamma \big(l+3/2\big)}r^l+\beta \frac{(\omega/2)^{-l-1/2}}{\Gamma \big(-l+1/2\big)}r^{-l-1}~.
\end{align}
We have used the expansion of the Bessel function for small values of r, i.e., $J_{b}(r)=\{1/\Gamma(b+1)\}\left(r/2\right)^b$, to arrive at the above result.
Along similar lines we can  determine the behaviour of the solution written down in \eqref{infty_soln} for large values of  $r$ as
\begin{align}  \label{soln_infty_large}
R\sim\frac{1}{r}\sqrt{\frac{2}{\pi \omega}}\left[\alpha \text{sin}(\omega r-l\pi/2)+\beta \text{cos}(\omega r+l\pi/2)\right]~.
\end{align}

\subsection{Near-horizon solution}
To write the radial wave equation \eqref{wave_eq_hd} in the near-horizon limit defined as $\omega (r-r_+)\ll 1$, we define a new variable 
\begin{align} \label{z}
z=\frac{r-r_{+}}{r-r_-};    
\qquad
\Delta\partial_r=(r_{+}-r_-)z\partial_z~.
\end{align}
In terms of this new variable and using the near-horizon approximation appropriately, from \eqref{wave_eq_hd} we obtain,
\begin{align} \label{wave_eq_near}
z(1-z)\partial^2_zR+(1-z)\partial_zR+\bigg[{(ma-2Mr_+\omega)^2\over z(r_+-r_-)^2}-{(ma-2Mr_-\omega)^2\over (r_+-r_-)^2}
-{l(l+1) \over1-z}\bigg]R=0.
\end{align}
To write the above equation as the standard differential equation for a hypergeometric function we redefine the wavefunction $R$ as
\begin{align}
R=z^{i \sigma}(1-z)^{l+1}F,
\end{align}
where 
\begin{align}
&\sigma={\omega-m\Omega_H\over4 \pi T_H} \label{sigma},
\\
&\sigma'={2\omega Mr_--ma\over r_+-r_- },
\end{align}
 we define $\sigma'$ for later convenience.
 Using this, \eqref{wave_eq_near} becomes
 \begin{align}
 z(1-z)\partial^2_zF+\big(1+i2\sigma-(1+2(l+1)+i2\sigma)z\big)\partial_zF
 -\big((l+1+i\sigma+i\sigma')(l+1+i\sigma-i\sigma')\big)F=0.
 \end{align}
The solution of the above equation is given in terms of hypergeometric functions as
\begin{align} \label{sol_near}
R=Az^{-i\sigma}(1-z)^{l+1}F[a-c+1,b-c+1,2-c,z]
+Bz^{i\sigma}(1-z)^{l+1}F[a,b,c,z],
\end{align}
where $a=1+l+i\sigma-i\sigma',b=l+1+i\sigma+i\sigma',c=1+2i\sigma$. 
For small values of $z$  i.e near the reflective membrane this solution can be expanded as   
\begin{align} \label{near_near}
R\sim Az^{-i\sigma}+Bz^{i\sigma}=Ae^{-i(\omega-m\Omega_H){r_+\over 2M}r^*}+Be^{i(\omega-m\Omega_H){r_+\over 2M}r^*},
\end{align}
where the tortoise coordinate $r^*$ is defined as
\begin{align} \label{tor}
r^*=\int {r^2+a^2\over(r-r_+)(r-r_-)}dr.
\end{align}
From \eqref{near_near} we can interpret the first part as the outgoing wave and the second part multiplied with B as the ingoing wave. The large r limit of the above solution \eqref{sol_near} is relevant in the intermediate matching region and it is given as

\begin{align}
\label{soln_near_large}
R\sim
\bigg(&\frac{r}{r_{+}-r_-} \bigg)^l~\Gamma(2l+1)\bigg[ A\frac{\Gamma(1-2Q)}{\Gamma(1+l-Q-Q')\Gamma(l+1-Q+Q')}+B\frac{\Gamma(1+2Q)}{\Gamma(l+1+Q+Q')\Gamma(l+1+Q-Q')}\bigg]
\nonumber\\
&+\bigg(\frac{r}{r_{+}-r_-} \bigg)^{-l-1}~\Gamma(-2l-1)\bigg[ A\frac{\Gamma(1-2Q)}{\Gamma(-l-Q-Q')\Gamma(-l-Q+Q')}+B\frac{\Gamma(1+2Q)}{\Gamma(-l+Q+Q')\Gamma(-l+Q-Q')}\bigg], \nonumber\\
\end{align}
where we defined $Q=i\sigma$ and $Q'=i\sigma'$. 
\subsection{Matching region}
Assuming a reflective membrane, with reflectivity $\mathcal{R}$, at $z_0$ we can write the boundary condition at the horizon as
\begin{align} \label{boundarycon}
{B\over A}z_0^{2i\sigma}=\mathcal{R}e^{i\pi\delta}
\end{align} 
Now matching \eqref{soln_infty_small} and \eqref{soln_near_large} in the intermediate region $(1/\omega\gg r\gg M)$ and using the boundary condition \eqref{boundarycon} we get two independent equations that will be used for computation of the flux
\begin{align} \label{amplitude_rel}
&{\alpha\over A}=\bigg({\omega\over 2}\bigg)^{-l-{1\over 2}}{\Gamma(l+{3\over 2})\Gamma(2l+1)\over (r_+-r_-)^l}\bigg[{\Gamma(1-2Q)\over\Gamma(1+l-Q-Q')\Gamma(l+1-Q+Q')}+z_0^{-2i\sigma}\mathcal{R}e^{i\pi\delta}{\Gamma(1+2Q)\over\Gamma(l+1+Q+Q')\Gamma(l+1+Q-Q')}\bigg]
\\
&{\beta\over A}=\bigg({\omega \over 2}\bigg)^{l+{1\over 2}}{\Gamma(-l+{1\over 2})\Gamma(-2l-1)\over (r_+-r_-)^{-l-1}}\bigg[{\Gamma(1-2Q)\over\Gamma(-l-Q-Q')\Gamma(-l-Q+Q')}+z_0^{-2i\sigma}\mathcal{R}e^{i\pi\delta}{\Gamma(1+2Q)\over\Gamma(-l+Q+Q')\Gamma(-l+Q-Q')}\bigg] \label{beta/A}
\end{align}
One must note that the result doesnot depend on the arbitrary choice of the matching surface. If $r_{match}$ is the matching surface the solution of the wave equation must be invariant under any change in $r_{match}$ for the procedure to work. Changing $r_{match}$ would also change the redshift factor at the matching surface which would indicate a local change in scale. This hints at a local conformal symmetry of the solutions. 

To obtain the QNM spectrum of the ECO we must use an outgoing boundary condition at infinity. 
This would mean, in the wavefunction of the far region given in \eqref{soln_infty_large} there would be no incoming wave from infinity, giving us the relation $\beta=-i\alpha e^{i\pi l}$.
As we discussed before, for a Kerr-like ECO the near-horizon boundary condition is modified and is given by \eqref{boundarycon}. Using these boundary conditions, matching \eqref{soln_near_large} and \eqref{soln_infty_small} to eliminate the unknown constants($A,B,\alpha,\beta$) we get
\begin{align} \label{matching_z_0}
z_0^{2i \sigma}\prod_{n=1}^{l}\bigg[ \frac{n+2 i\sigma}{n-2i\sigma}\bigg]
\bigg[\frac{1+2L(r_+-r_-)^{2l+1}\sigma\omega^{2l+1}\prod_{n=1}^{l}(n^2+4 \sigma^2)}{1-2L(r_+-r_-)^{2l+1}\sigma\omega^{2l+1}\prod_{n=1}^{l}(n^2+4\sigma^2)}  \bigg]
=-\mathcal{R}e^{i\pi\delta},
\end{align}where $L$ in the above equation is defined as
\begin{align}
L\equiv \frac{\pi \{\Gamma(l+1)\}^2}{2^{2l+2}\Gamma(l+3/2)\Gamma(2l+2)\Gamma(2l+1)\Gamma(l+1/2)}.
\end{align}Usually one needs to resort to various numerical techniques for solving \eqref{matching_z_0} but we can solve it approximately in the low frequency limit to obtain the quasi normal frequencies. Near the superradiant bound $\omega=m\Omega_H$ we can assume $\sigma \ll 1$, which is same as the low frequency limit $M\omega \ll 1$, we are working in. In this limit, from \eqref{matching_z_0} and using the tortoise coordinates we get
\begin{align} \label{match_low_freq}
z_0^{2i \sigma}=e^{2i\sigma r_0^*(r_+-r_-)/(2Mr_+)}=-\mathcal{R}e^{i\pi\delta}.
\end{align}
This can be solved to obtain the QNM spectrum as,
\begin{align} \label{qnf}
\omega_n-m\Omega_H={\pi(2n+1+\delta)\over 2r^*_0}-i{\ln\mathcal{R}\over 2r_{0}^{*}}.
\end{align}

 \section{Two-dimensional conformal field theory on a torus} 
 In AdS/CFT the  thermal state of a black hole corresponds to the thermal state of the boundary CFT. Perturbations in the bulk corresponds to perturbations in the boundary thermal field theory with operator $\mathcal{O}_{(h,\bar h)}$. The time evolution of the perturbation and the evolution of the system is described in terms of the linear response theory. According to this theory all the relevant information about the evolution of the perturbation and the relaxation of the system is contained in the retarded two-point function. 
Now our aim in this paper is to understand the quantum modification to the near-horizon geometry of a Kerr spacetime within the context of the Kerr/CFT correspondence. To do this we need a better understanding of the dual two-dimensional CFT living in the near-horizon region of the Kerr spacetime as proposed by the correspondence and compute the appropriate thermal correlation function.  In this appendix we first review some basic features of a two-dimensional conformal  field theory and discuss about the  two-point function of a  CFT living on a torus.

We start with a field theory on an Euclidean plane with coordinates $x_1$ and $x_2$ and define a set of complex coordinates as 
\begin{align}
w=x_1+ix_2 \quad \bar w=x_1-ix_2.
\end{align}
A field $\phi(w,\bar w)$ is called a primary field of conformal dimension $(h,\bar h)$ if it transforms under a conformal transformation $w\to f(t_R), \bar w\to f(t_L) $ as

\begin{align}
\phi(t_R,t_L)=f'(w)^h\bar f'(\bar w)^{\bar h} \phi(f(w),\bar f(\bar w)).
\end{align}
Based on the transformation properties of the local CFT operators, it is ensured that the correlation functions will transform covariantly under the conformal group. This would determine and fix the form of the two-point function of CFT operator $O(w,\bar{w})$as

\begin{align} \label{2pointplane}
\langle O_1(w_1,\bar{w_1})O_2(w_2,\bar w_2)\rangle=\frac{C_{12}}{(w_1-w_2)^{2h_R}(\bar w_1-\bar w_2)^{2 h_L}},
\end{align}
where $C_{12}$ is a constant and $h_R, h_L$ is related to the conformal dimension of the operators $O_1,O_2$.


\subsection{Two-point function on the torus}\label{2point}
A two-dimensional torus can be described in terms of  coordinates $(x,\tau_E)$ with periodicity $(x,\tau_E)\to (x+L,t_E+T^{-1})$. This torus  is characterised  by the modular parameter  $\tau=iT^{-1}/L$ and the complex holomorphic coordinate $t_R=x+i\tau_E$, having periodicity $t_R\to t_R+nL+imT^{-1}$, for $n,m \in \mathbb{Z}$ .  The anti-holomorphic coordinate can be defined as $ t_L=x-i\tau_E$. The modular transformation, characterising the modular parameter and the holomorphic coordinate is given as 
\begin{align}
\tau^{'}={\mathbf{a} \tau+\mathbf{b}\over \mathbf{c}\tau+\mathbf{d}},\quad t_{R}^{'}={t_{R}\over \mathbf{c}\tau+\mathbf{d}}, \quad \mathbf{ad-bc}=1.
\end{align}
 For fields having conformal weights $h_L,h_R$ the correlation function is given by \eqref{2pointplane}.
 The relation between $w$ and the torus coordinate $t_R$ can be defined as $w=e^{it_R/(\mathbf{c}\tau+\mathbf{d})}$. In terms of the torus coordinates we can define the two-point function (after performing the sum over images) as

\begin{align}
\langle O_1(t_R, t_L)&O_2(0,0)\rangle_{SL(2,Z)}=
\nonumber\\
&\sum_{n\in \mathbb{Z}}\left|2(\mathbf{c}\tau+\mathbf{d}) \text{sin}\pi\bigg[{t_R+ nL(\mathbf{a}\tau+\mathbf{b})\over  L(\mathbf{c}\tau+\mathbf{d})}\bigg]\right|^{-2h_R}
\left|2(\mathbf{c}\bar\tau+\mathbf{d}) \text{sin}\pi\bigg[{t_L+n L(\mathbf{a}\bar\tau+\mathbf{b})\over  L(\mathbf{c}\bar\tau+\mathbf{d})}\bigg]\right|^{-2h_L},
\end{align}
where $\tau=iT_{R}^{-1}/L, \bar\tau=iT_{L}^{-1}/L$

We must note that, according to the sum over geometries prescription of \cite{Maldacena:2001kr} if we want to compute the two-point function on the torus one must sum over the $SL(2,\mathbb{Z})/\mathbb{Z}$ family with an appropriate weight factor which usually depends on the action\cite{Dijkgraaf:2000fq}. Here we are interested in computing the two-point function of a particular geometric configuration corresponding to a specific set of values for the group elements and obtain the QNM spectrum form the poles of the two-point function. Hence we are ignoring the weight factor in front that depends on the action as it would not contribute to the pole.  Now for computing the real time correlation function, we perform the analytic continuation $\tau_E\to it$, hence $t_R=x-t$ and $t_L=x+t$, this will give us the two-point function as

\begin{align} \label{2pointtorus}
\langle O_1(t_R, t_L)O_2(0,0)\rangle_{SL(2,Z)}=\sum_{ n\in \mathbb{Z}}\left[{L^2\over \mathbf{c}^2T_{R}^{-2}+\mathbf{d}^2L^2}\right]^{2h_R}\left[{L^2\over \mathbf{c}^2T_{L}^{-2}+\mathbf{d}^2L^2}\right]^{2h_L}{(\pi /L)^{2h_R}(\pi /L)^{2h_L}\over[\text{sinh}\pi(x_R)]^{2h_R}[\text{sinh}\pi(x_L)]^{2h_L}},
\end{align}
where 
\begin{align}
x_{R,L}={T_{R,L}\over \mathbf{c}^2T_{R,L}^{-2}+\mathbf{d}^{2}L^{2}}(\mathbf{c} t_{R,L}\pm nL)-i{L\over \mathbf{c}^2T_{R,L}^{-2}+\mathbf{d}^2L^2}(\mathbf{d} \,t_{R,L}+n{\mathbf{ac}T_{R,L}^{-2}+\mathbf{bd}L^2\over L}).
\end{align}

For a black hole choosing the  $SL(2,\mathbb{Z})$  parameter as $(\mathbf{a}=0,\mathbf{b}=1,\mathbf{c}=-1,\mathbf{d}=0)$ \footnote{ In the case of $AdS_3/CFT_2$, for the $SL(2,Z)$ parameters we know the choice $(\mathbf{a}=1, \mathbf{b}=\mathbf{c}=0,\mathbf{d}=1) $ corresponds to thermal $AdS$ while $(\mathbf{a}=0, \mathbf{b}=1,\mathbf{c}=-1,\mathbf{d=0}) $ corresponds to the BTZ black hole. } we obtain the two-point function on the tours as \cite{Chekhov:1999uk,Chekhov:2004hu,Birmingham:2002ph,Solodukhin:2004rv}

\begin{align} \label{2pointtorus1}
\langle\mathcal{O}(t_R,t_L)\mathcal{O}(0,0)\rangle_{torus} =
\sum_{n\in \mathbb{Z}}{(\pi T_R)^{2h_R}(\pi T_L)^{2h_L} \over[\text{Sinh}(\pi T_R(t_R+nL))]^{2h_R}[\text{Sinh}(\pi T_L(t_L+nL))]^{2h_L}}.
\end{align}
The torus two-point function is doubly periodic as expected and it can be  expressed in terms of  basic elliptic functions (double periodic Eisenstein Weierstrass series), which is given as \cite{Barrella:2013wja,Chekhov:2004hu,Nikolov:2005df,Nikolov:2004yg} 
\begin{align}
p_{2h}(z|T)=
\sum_{m,n\neq 0,0}{1\over (z+mL+in/T)^{2h}}.
\end{align}
where again the two periods $(L,i/T)$ can be identified with the coordinates on the torus $t\sim t+L\mathbb{Z}+{i\mathbb{Z}\over T}$. For $h$=1 in \eqref{2pointtorus}, using Euler's formula we can write the 2 point function as \cite{Chekhov:2004hu,Nikolov:2004yg}

\begin{align} \label{weie}
\langle\mathcal{O}(t_R,t_L)\mathcal{O}(0,0)\rangle_{torus}=
\sum_{m,n\neq 0,0}{1\over (t_R+m_RL+in/T_R)^{2}}{1\over (t_L+m_LL+in/T_L)^{2}}
 =p_{2h}(t_R|T_R)p_{2h}(t_L|T_L).
\end{align}

For $h>1$ \eqref{2pointtorus} can be written as a series of higher order Eisentein functions \cite{Nikolov:2005df}.

\subsection{The Cardy formula} \label{cardy}

Using two linearly independent lattice vectors (periods of the lattice $\omega_1 ~\text{and} ~\omega_2$) on the complex plane and identifying points that differ by an integer combination of $\omega_1, \omega_2$, we can define a torus. The CFT defined on the torus is independent of the overall scale of the lattice and the modular
parameter, $\tau=\omega_2/\omega_1$ is the only relevant parameter (we will take the modular parameter as purely imaginary and given by $\tau=\beta/2\pi L$ ). 

We can write the partition function $Z$ in terms of the Hamiltonian (H) and momentum (P) of the theory.  A translation operator that translates the system parallel to the period $\omega_2$ over a distance $a$ is given as
\begin{align}
e^{-{a\over \omega_2}[H ~\text{Im} \omega_2-iP~\text{Re} \omega_2]}.
\end{align}
Defining  $a$ as the lattice spacing, this translation operator would take us from one row of the lattice to the next one (parallel to $\omega_2$). The partition function can be written as 
\begin{align} \label{partition}
Z(\omega_1, \omega_2)=\text{Tr} ~e^{-[H ~\text{Im} \omega_2-iP~\text{Re} \omega_2]}.
\end{align}
Expressing the operators $H$ and $P$ in terms of the Virasoro generators $(L_0,\bar{L_0)}$
\footnote{The torus can be constructed by gluing the ends of an infinite cylinder. If we define the circumference of the cylinder as $R$, the Hamiltonian and the momentum can be written as
\begin{align}
&H=2\pi/R(L_0+\bar{L_0}-c_L/24-c_R/24)
\\
&P=2\pi i/R(L_0-\bar{L}_0)
\end{align}$c_L,c_R$ are the central charges of the CFT as we defined before. } 
we can write \eqref{partition} as \begin{align}
Z(\tau, \bar{\tau})\sim \text{Tr} ~e^{2\pi i[\tau(L_0-c_L/24)-\bar{\tau}(\bar{L}_0-c_R/24)]}.
\end{align}
If we take $c_L=c_R=c$, the energy eigenvalue $L_0=E_L$ and focusing on the $\tau$ part, the density of state is given by the following contour integral:
\begin{align}
\rho(E)=\int_C d\tau Z(\tau)e^{-2\pi i\tau(E_L-c/24)}.
\end{align}
To compute the density of state, the trick is to use the modular invariance of the partition function $(Z(\tau)=Z(-1/\tau))$ to relate the low temperature and the high temperature behaviour. We can further use a saddle point approximation to compute the dominant contribution to the integral and obtain the  density of state as:
\begin{align}
\rho(E_L,E_R)\propto \text{const} \times \text{exp}\bigg[2\pi \sqrt{\frac{c_L}{6}\bigg( E_L-\frac{c_L}{24}\bigg)}+2\pi \sqrt{\frac{c_R}{6}\bigg( E_R-\frac{c_R}{24}\bigg)}\bigg].
\end{align}
One must note that this expression is valid in the limit $E\gg c/12$(high temperature limit). Taking the logarithm of the density of state(neglecting the constant in the front as it is small in the high temperature limit) we get the entropy as 
\begin{align}
S_{Cardy}=2\pi\sqrt{\frac{c_L}{6}\bigg(E_L-\frac{c_L}{24}\bigg)}+2\pi\sqrt{\frac{c_R}{6}\bigg(E_R-\frac{c_R}{24}\bigg)}.
\end{align}
This is the well known {\it Cardy formula} and we can see interestingly that it depends on the central charge and the energy eigenvalues. Further, it is possible to write the Cardy formula in terms of the right and left temperature for a canonical ensemble. The temperatures are defined as 
\begin{align}
{\partial S\over\partial E_L}={1\over T_L},
\\
{\partial S\over\partial E_R}={1\over T_R}.
\end{align}
In terms of the temperature the Cardy formula is given as 
\begin{align}
S_{Cardy}={\pi^2 \over 3}(c_L T_L+c_R T_R).
\end{align}
We must note that incase we lift the CFT to a covering space which is a n times longer circle than the original spatial length of the torus both the central charge and the temperature would get rescaled as $\tilde c=c/n, \tilde T=nT$ making the entropy invariant\cite{Balasubramanian:2014sra,Guica:2010ej}. 
Also, recently it was shown in \cite{Chen:2020nyh}, the central charge can be expressed in terms of the left and right temperature as $c_{L/R}\propto 1/(T_L+T_R)$ . Thus, if we consider a rescaled temperature by some constant factor, the combination $c_LT_L+c_RT_R$ remains unchanged making the entropy invariant under such rescaling.

\bibliography{echoes}
\bibliographystyle{./utphys1}

\end{document}